# A 100 GHz Wideband Reconfigurable Intelligent Surface Based on Orthogonal Polarization and Sub-Array Partitioning Concepts

Ruiqi Wang, Behrooz Makki, *Senior Member, IEEE*, and Atif Shamim, *Fellow, IEEE*

***Abstract*—The sub-terahertz (sub-THz) frequency band offers extremely large bandwidth and enables ultra-high data rates for future wireless applications. However, severe propagation loss and blockage from the environment significantly limits coverage at these frequencies. A solution to this problem can be Reconfigurable intelligent surfaces (RIS), which can dynamically shape the electromagnetic wave propagation in the wireless channel. Realizing RIS using standard Printed Circuit Board (PCB) technology is desirable due to its low cost and wide availability. However, such PCB-based implementations remain difficult due to fabrication resolution limitations, the lack of suitable RF switches, and the incompatibility between the large physical size of switches and the small unit-cell dimensions at these frequencies. Another issue is that conventional multi-resonance-based techniques, to achieve wideband performance, become less effective due to the strong parasitic effects introduced by switches at sub-THz frequencies. To alleviate these problems, a switch-aware EM design approach is required, where the unit-cell geometry and feeding topology are jointly optimized under switch constraints to accommodate hardware limitations. In this work, we address these limitations and demonstrate a wideband PCB-based RIS operating in the sub-THz band (center frequency 100 GHz) for the first time. A subarray partitioning architecture is adopted to make the layout feasible and reduce the number of switches, thereby lowering hardware cost and enabling scalable implementation. Meanwhile, an orthogonal polarization-based microstrip slot-coupled patch structure is employed to mitigate switch-induced parasitic effects and achieve a wideband response. The reconfigurability is achieved using AlGaAs SP3T switches integrated through bond wires, enabling a fast-switching speed of approximately 2 ns, however, the wire bonding has to be optimized using multiple parallel bond wires. In addition, the switches are positioned close to the RF transmission lines to maintain short and nearly vertical interconnections. For proof of the concept, a six-subarray RIS with 4×4 elements per subarray is designed for different beamforming angles, and the final 12×8 array is simulated, fabricated, and experimentally characterized. The measurement results demonstrate a gain enhancement of approximately 10 dB across 86–100 GHz band and about 5 dB across 100–106 GHz band, while maintaining a low power consumption of 0.165 W and reducing the overall hardware cost by employing only two switches, validating the effectiveness of the proposed wideband sub-THz RIS prototype.**

## I. INTRODUCTION

THE sub-terahertz (sub-THz) frequency band, typically spanning 90–300 GHz, has attracted increasing attention due to the availability of extremely large bandwidth [1], [2]. Such abundant spectrum resources enable ultra-high data rate wireless communication and support a wide range of bandwidth-intensive applications, such as holographic communication, immersive extended reality, high-resolution sensing, and intelligent machine-to-machine interaction [3], [4]. Compared with conventional frequency bands, sub-THz systems offer the potential for significantly enhanced capacity, sensing resolution, and data throughput. However, wireless communication and sensing at sub-THz frequencies encounters several critical challenges. Due to the very short wavelength, electromagnetic (EM) waves in this frequency band experience strong atmospheric absorption, and high sensitivity to blockage from obstacles such as buildings, vehicles, and human bodies. These propagation characteristics significantly limit the coverage range and reliability of sub-THz wireless links. Therefore, it becomes essential to use techniques that can dynamically control and manipulate the wireless propagation environment.

Reconfigurable intelligent surfaces (RIS) have recently emerged as a promising solution for enhancing wireless communication and sensing performance by intelligently shaping the electromagnetic wave propagation [5], [6]. In this context, RISs can provide programmable EM environments for sub-THz systems, where controllable reflections can help improve signal coverage, illumination, and environment perception. One of the main motivations for RIS, compared with other alternatives such as network-controlled repeaters, is its low-cost implementation. Therefore, to support practical deployment, the per-node cost of RIS needs to be kept low. However, even with low hardware cost, RIS may still face different implementation and regulatory considerations, in

Manuscript received 24th May 2026. This work was supported by Ericsson Research under Grant ORA#6089. (*Corresponding author: Ruiqi Wang.*)

R. Wang and A. Shamim are with Computer, Electrical and Mathematical Sciences and Engineering Division, King Abdullah University of Science and Technology (KAUST), Thuwal, 23955-6900, Saudi Arabia (email: ruiqi.wang.1@kaust.edu.sa, atif.shamim@kaust.edu.sa).

Behrooz Makki is with Ericsson Research, Ericsson, 417 56 Gothenburg, Sweden (e-mail: behrooz.makki@ericsson.com).

cellular networks, as discussed in [7]. Nevertheless, RIS may also be useful in other non-cellular network applications. An RIS typically consists of a large number of reconfigurable scattering elements whose reflection responses can be dynamically controlled. By properly adjusting the reflection phase and amplitude of each element, RIS can redirect electromagnetic waves toward desired directions, thereby improving signal coverage, mitigating blockage, and enhancing communication reliability. As a result, RIS technology has been investigated as a potential enabling hardware platform for next-generation wireless communication systems [8][1].

Despite the significant potential of RIS technology, realizing practical RIS prototypes operating at sub-THz frequencies remains a significant challenge [9], [10]. Typically, high fabrication resolution with fast switching speeds can be achieved through CMOS process. However, they usually involve extremely high fabrication costs and limited aperture sizes [11], [12]. Therefore, low-cost and scalable implementations based on commercially available printed circuit board (PCB) technology are desirable. However, the implementation of RIS using PCB technology at sub-THz frequencies is particularly challenging.

First, the extremely short wavelength in the sub-THz band leads to very small unit-cell dimensions, imposing stringent requirements on fabrication precision. In our previous studies [13]–[15], a series of millimeter-wave (mm-Wave) RIS prototypes operating around 30 GHz were developed. Even at these frequencies, the unit-cell size already approached the practical fabrication limit, indicating that further scaling toward sub-THz frequencies becomes even more challenging. In addition, the availability of suitable RF switches that are compatible with standard PCB processes at these frequencies is limited, which restricts the implementation of reconfigurable functionality in practical RIS systems. Moreover, the physical size of existing switches is relatively large compared with the small feature size of RIS unit cells at sub-THz frequencies, making direct integration of the switches with the EM structures very challenging. Furthermore, assigning one switch per unit cell leads to high hardware cost and power consumption, which limits the scalability of large-aperture RIS implementations. This challenge calls for a switch-aware EM design strategy, in which the unit-cell geometry, feeding topology, and switch placement are jointly optimized to accommodate the physical size and biasing requirements of the switching components while preserving the desired EM performance. In addition, conventional multi-resonance-based wideband techniques become less effective at sub-THz frequencies due to the strong parasitic effects introduced by switches, which significantly distort the reflection phase response. As a result, developing a low-cost, scalable, and PCB-compatible RIS operating at sub-THz frequencies remains an open problem. To the best of the authors' knowledge, no experimentally demonstrated wideband PCB-based RIS operating in this frequency range (center frequency 100 GHz) has been previously reported.

In this work, a wideband PCB-based RIS design operating in the sub-THz band is developed to address the practical challenges of implementing RIS at high frequencies. To address the switch size constraint and reduce the number of required switches for scalable implementation, a subarray-based partitioning architecture is introduced, where multiple RIS elements are grouped to achieve controllable reflection responses for different beamforming directions. Meanwhile, an orthogonal polarization scheme based on an aperture-coupled patch structure is employed to mitigate switch-induced parasitic effects and enable wideband performance. The proposed design adopts a multilayer PCB platform combined with high-speed AlGaAs switches to enable dynamic beam manipulation while maintaining a relatively low fabrication cost. The fabricated prototype demonstrates effective beam manipulation capability and provides ~10 dB gain enhancement over a wide frequency range around 100 GHz. These results verify the feasibility of realizing low-cost [2], scalable, and wideband RIS implementations in the sub-THz regime using a PCB-based platform.

## II. RIS Unit Cell Design

### A. Choice of 100 GHz Switch

To realize the reconfigurable functionality of the sub-THz RIS, the type of radio-frequency (RF) switch must be determined prior to the EM design. For mm-Wave RIS implementations, a wide range of commercial PIN diode switches are available. A representative example is the MA4AGFCP910 [16], which provides acceptable RF performance with insertion loss below 0.8 dB in the ON state and isolation higher than 10 dB in the OFF state up to 30 GHz. In addition, several circuit techniques have been used to enhance the EM performance of RIS unit cells incorporating switches, such as slotline ring topologies [17] and coupling-enhancement structures [18]. Therefore, for RIS designs operating in the mm-Wave band, the availability of suitable switching components generally does not pose a major limitation.

However, when the operating frequency increases to the sub-THz region around 100 GHz, the selection of commercially available RF switches becomes extremely limited, as summarized in Table I. Although several waveguide-based switches, such as SP4T [19] and SPDT [20] configurations, can operate within the W-band, they typically suffer from bulky physical dimensions, high insertion loss, and high cost. More importantly, their low integration compatibility makes them unsuitable for large-scale RIS implementations based on PCB platforms. In contrast, MMIC-based switches offer a much more compact form factor and better compatibility with planar fabrication processes. However, only a very limited number of such devices can operate close to 100 GHz while maintaining acceptable RF performance. Among them, the MASW-011111

[1] RIS is not part of the 5G NR standardization. Moreover, 3GPP has not yet reached an agreement on including RIS in 6G study items. The practical implications of RIS in cellular networks are discussed in [7].

[2] In this paper, cost refers to the per-node cost of the RIS. In practice, the total cost of a RIS-assisted network depends on various parameters, including the per-node cost, node density in the deployment area, site rental cost, and other deployment-related expenses.

TABLE I. COMPARISON OF COMMERCIAL W-BAND RF SWITCHES

| Name | Switch type | Switch function | Frequency range (GHz) | Size ($mm^3$) | Insertion loss (dB) | Isolation (dB) | Cost | Integration compatibility |
|---|---|---|---|---|---|---|---|---|
| SK4-7531148030-1010-R1-M [18] | Waveguide | SP4T | 75–110 | 40.6 × 43.2 × 24.1 | 8 | 30 | High | Low |
| PE71S9004 [19] | Waveguide | SPDT | 75–110 | 71.9 × 30.5 × 19.1 | 5.5–7 | 20 | High | Low |
| MASW-011111 [20] | MMIC Die | SPDT | 65–100 | 1.33 × 1.06 × 0.1 | 0.8 | 25 | Medium | High |
| MASW-011029 [21] | MMIC Die | SP3T | 60–110 | 1.6 × 1.13 × 0.1 | 1.3 | 33 | Medium | High |

[21] and MASW-011029 [22] represent two typical commercially available solutions. Considering both cost and availability, the MASW-011029 switch is selected in this work to realize the reconfigurable functionality of the RIS. The MASW-011029 is a wideband single-pole triple-throw (SP3T) switch that provides a fast switching speed on the order of 2 ns, making it suitable for high-speed reconfigurable EM structures. Its physical layout is illustrated in Fig. 1. The device includes one RF input port and three RF output ports, where each path can be independently controlled via biasing to select the desired transmission channel.

From an RIS hardware design perspective, several practical considerations arise when integrating this type of switch into a PCB-based architecture. First, the physical dimensions of the switch are approximately 1.6 mm × 1.13 mm, corresponding to about 0.53λ × 0.38λ at 100 GHz. Compared with the extremely small wavelength at sub-THz frequencies, this footprint is relatively large for an RIS unit cell, which is typically a half-wavelength. As a result, the EM design of the RIS cannot be performed independently of the switching component. Instead, the unit-cell geometry and feeding topology need to be co-designed with the switch placement to accommodate the physical size of the device while maintaining acceptable radiation and impedance characteristics. Overall, the limited availability and relatively large physical size of commercially available sub-THz switches impose significant constraints on the RIS architecture. Consequently, the EM design of the proposed RIS needs to be carried out while accounting for the physical dimensions, bias requirements, and RF performance of the selected switch. In addition, the MASW-011029 is provided as a bare die rather than a packaged component. The backside metal of the die needs to be connected to the RF and DC ground, and therefore conventional chip-level assembly techniques such as solder die attach or conductive epoxy bonding are required. After die attachment, RF and DC connections are established using gold wire bonding. The bonding inductance and the routing of the RF traces need to be carefully controlled to avoid degrading the RF performance, particularly at frequencies close to 100 GHz.

### B. Unit Cell Design with Orthogonal Polarization

As illustrated in Section A, the RIS design is constrained by the switch due to its relatively large footprint. The parasitic effects introduced by the switch, including equivalent phase delay and capacitive loading, significantly degrade the EM performance of the unit cell, particularly the reflection phase response. Consequently, the multi-resonance techniques adopted in our previous works [13], [14] are no longer suitable for sub-THz unit cell design. To address this issue, a unit cell design method based on current reversal is adopted, as conceptually illustrated in Fig. 2. Specifically, the radiation element first captures the incident field and transfers it to the RF input port of the switch. Then, two opposite current transmission paths are selectively controlled by the two RF output ports of the switch, and the signal is routed back to the radiation element with cross-polarized reflection. These two opposite current paths are equivalent to a physical rotation of the unit cell, thereby enabling a natural 180° phase difference. This design approach provides two key advantages. First, the switch parasitic effects, including phase delay and capacitive loading, are effectively cancelled between the two differential current paths from the two switch output ports. Meanwhile, the orthogonal polarization between the incident and reflected waves provides intrinsic EM isolation.

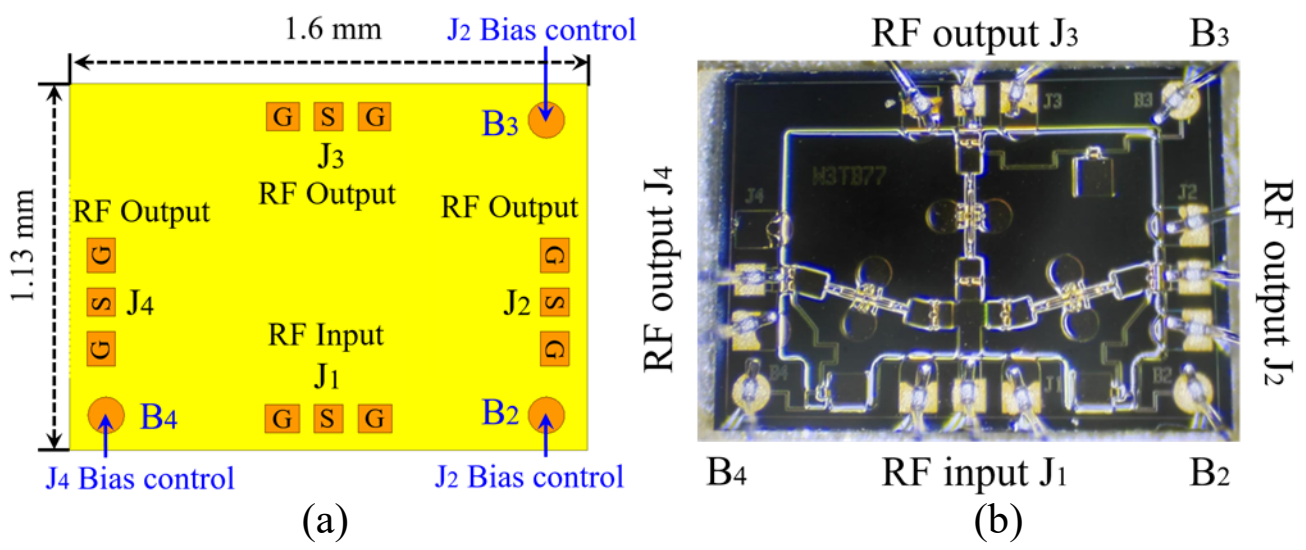

(a) (b)

Fig. 1. The wideband 75 - 100 GHz AlGaAs SP3T PIN Diode Switch. (a) Layout. (b) Die.

Based on the switch-aware orthogonal polarization structure with current reversal, the design evolution of the RIS unit cell is illustrated in Fig. 3. The initial configuration adopts a slot-coupled patch structure, as shown in Fig. 3(a). In this design, a bow-tie shaped slot is employed to couple with the patch in order to achieve wideband performance. Meanwhile, the ground plane is introduced to suppress the back radiation. More importantly, this structure is compatible with the integration of the selected switch and the required bond-wire interconnection. Subsequently, the physical unit rotation technique is incorporated into the unit cell design, as illustrated in Fig. 3(b). By electrically rotating the unit cell using two current paths, the surface current direction is reversed, which enables a 180° phase difference between two reconfigurable states. Considering the relatively large footprint of the switch and the resonance stability, two neighboring unit cells are combined to form a single functional element, as illustrated in Fig. 3(c). In this configuration, the incident power is first combined and then split into two controllable directions. Finally, the integration of the switch into the unit cell is presented in Fig. 3(d). The switch is carefully positioned away from the slot region to avoid disturbing the resonance behavior, while vias are employed to connect the ground layer to the slot structure, providing a common ground reference for the RF and DC paths.

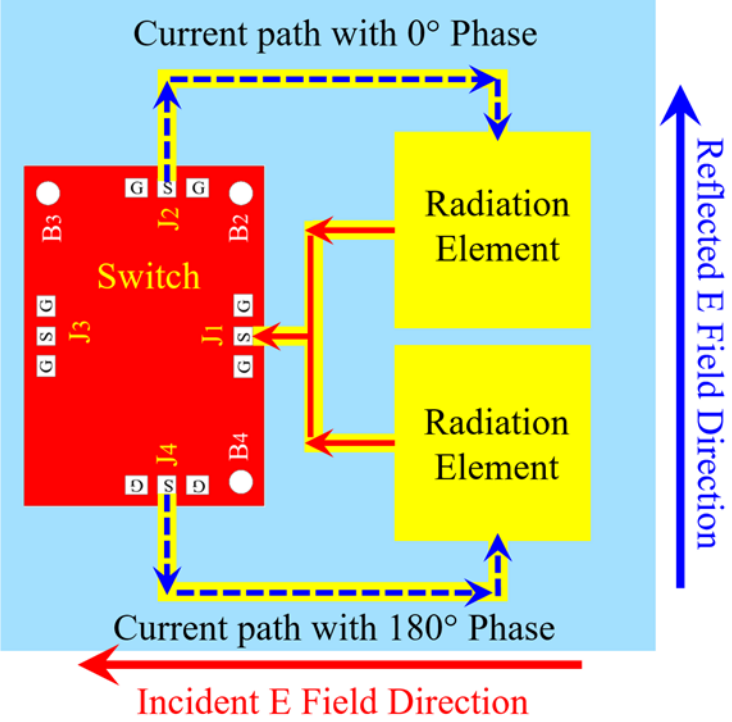

Fig. 2. Conceptual illustration of the current-reversal-based phase control mechanism.

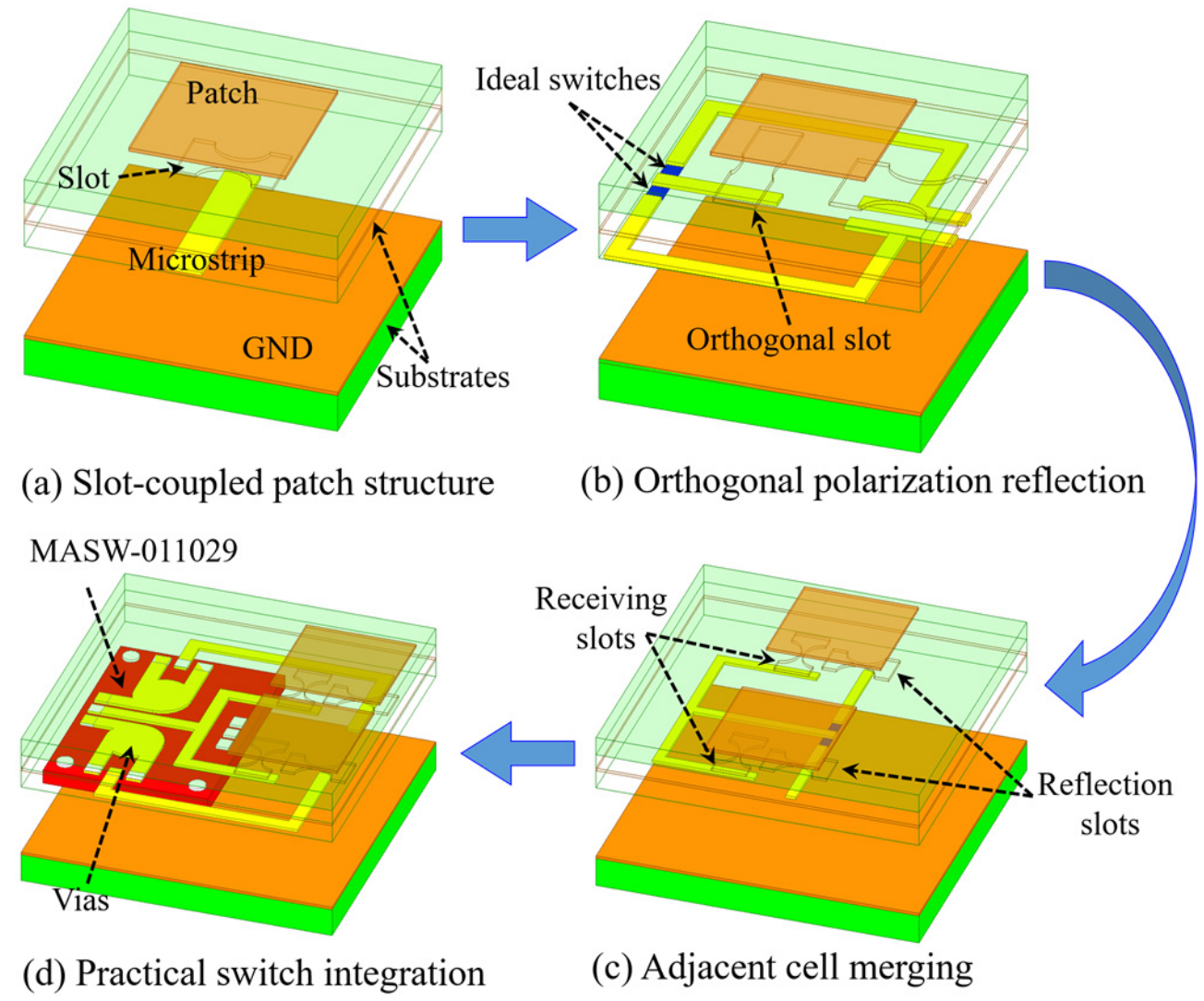

Fig. 3. The RIS unit design evolution.

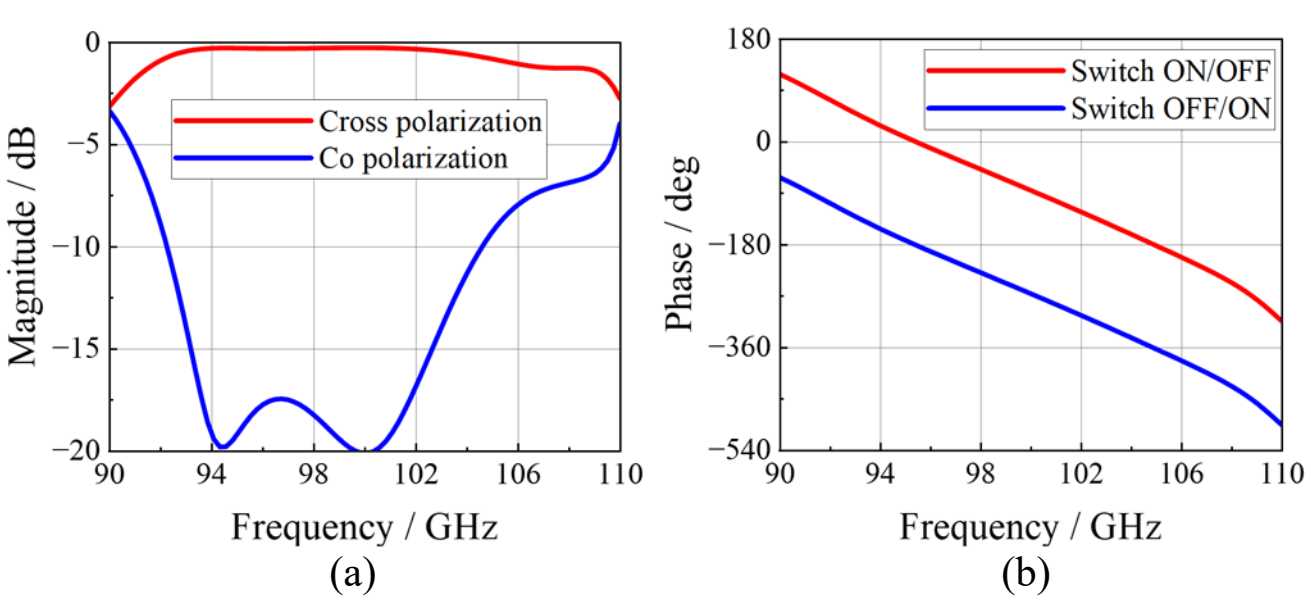

Fig. 4. RIS unit cell performance. (a) Reflection magnitude for co- and cross-polarizations. (b) Reflection phases for two cross-polarized reconfigurable states.

The simulated RIS unit cell performance is demonstrated in Fig. 4, which validates the effectiveness of the design strategies introduced in Fig. 3. From the results, a wideband reflection response can be observed. Specifically, the cross-polarization reflection magnitude is greater than −2 dB across most of the band from 90.9 to 109.6 GHz. Meanwhile, the co-polarization level is below −10 dB from 92.2 to 104.7 GHz, indicating that the incident co-polarized wave is effectively converted into the cross-polarization over a wide bandwidth. The phase difference between the two states is maintained around 180° ± 2° within the band of interest. This stable phase response further confirms the suitability of the proposed unit cell for phase-controlled wave manipulation.

## III. RIS Array Design

### A. Subarray Partitioning Concept

Based on the unit-cell structure shown in Fig. 3(d), the initial RIS array configuration is constructed by periodically arranging the unit cell, as illustrated in Fig. 5(a). In this configuration, each unit cell is directly integrated with one RF switch in order to provide independent control. However, such a dense-switch architecture introduces several practical challenges when operating at sub-THz frequencies. First, from an RF implementation perspective, the extremely dense distribution of switches makes wire bonding highly difficult, since multiple bond wires need to be placed within a very limited area while maintaining controlled inductance and avoiding mutual coupling between adjacent interconnections. Second, the physical footprint of the switch is relatively large compared with the wavelength at around 100 GHz. Consequently, the resulting unit-cell dimension (~0.7λ) becomes larger than half wavelength, which limits the angular stability of the RIS response. As demonstrated in Fig. 6, the reflection performance gradually degrades as the incident angle increases, and a rapid deterioration is observed when the angle exceeds approximately 10°, indicating limited angular stability of the RIS unit.

In addition to these RF limitations, several practical circuit-level issues also arise. One major concern is the routing of the DC biasing network. When a switching device is assigned to every unit cell, the available routing space for the DC biasing lines becomes highly limited, making it difficult to implement a scalable biasing network without introducing additional parasitic effects. Furthermore, the power consumption and hardware cost become significant when the array size increases. For example, a 20×20 RIS array would require 400 switches under the unit-level control architecture. For the MASW-011029 device, each isolation path requires approximately 10 mA bias current under a +5 V control voltage. Since two RF paths remain in the isolation state for an SP3T switch, the total DC current per switch is approximately 20 mA, corresponding to about 100 mW power consumption. Consequently, the total DC power consumption of a 400-switch RIS array would reach approximately 40 W, which introduces a considerable power burden for practical implementations. In addition, the overall hardware cost of such an architecture can become prohibitively high for large-scale RIS implementations.

To address these challenges, the RIS architecture is further evolved into the subarray-based configuration shown in Fig. 5(b). In this approach, multiple unit cells are grouped together to form a controllable subarray, where a single switching device controls the EM response of the entire element cluster. This subarray-based architecture significantly reduces the required number of switches, which not only simplifies the wire-bonding process but also alleviates the routing complexity of the DC biasing network. Moreover, the larger effective aperture of each subarray improves the angular stability of the RIS response while maintaining practical fabrication feasibility. As a result, the proposed architecture provides a more scalable and cost-effective solution for implementing wideband RIS systems in the sub-THz frequency band. For proof of concept, three distinct beams generated by different subarrays are designed towards −30°, 0°, and 30° directions, which are controlled by the three output RF paths of the MASW-011029 switch. The simulated radiation patterns of these subarrays under different beamforming directions are illustrated in Fig. 7. It can be observed that the three beams effectively cover the angular range from −45° to 45°, where the peak gains for the three beams are 14.3 dBi, 17.2 dBi, and 15.2 dBi, respectively. These results confirm the effectiveness of the proposed subarray-based architecture in achieving consistent and controllable beam switching performance.

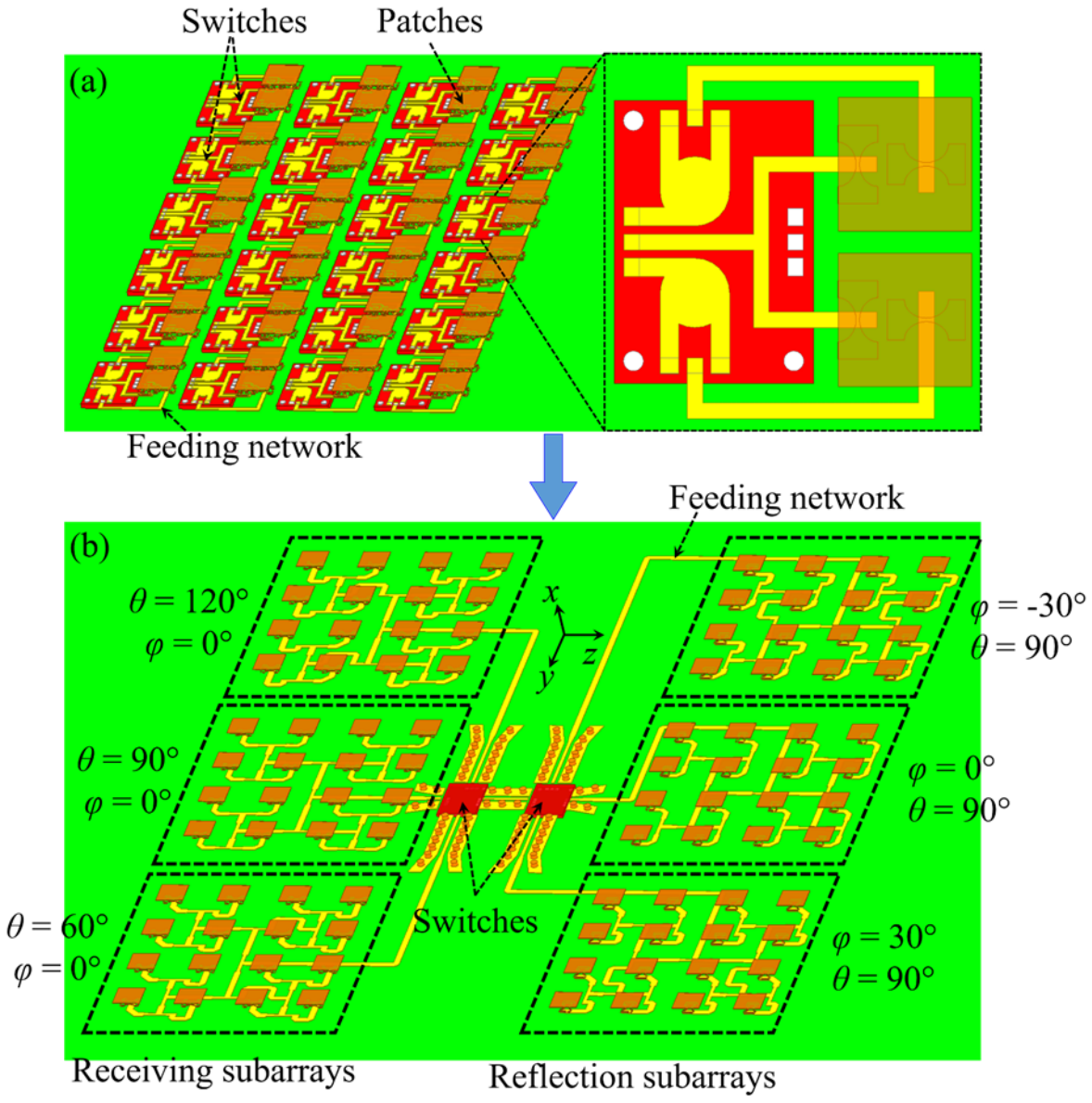

Fig. 5. The RIS array evolution.

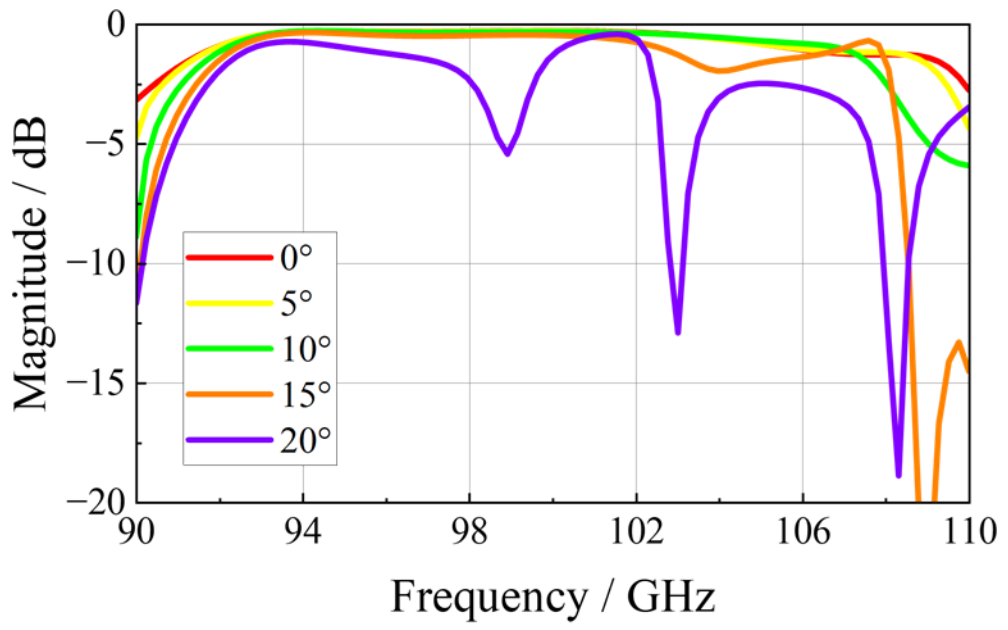

Fig. 6. Angular stability of the RIS unit.

Utilizing the subarray-based architecture discussed above, the overall RIS structure is implemented using a multilayer PCB platform, as illustrated in Fig. 8. The multilayer configuration provides sufficient routing space for both the RF feeding network and the DC biasing lines at sub-THz frequencies. As shown in Fig. 8(a), the radiating patch layer is implemented on a Rogers RT/duroid 5880 substrate to ensure low permittivity (~2.2) and dielectric loss (~0.002) for wide bandwidth for sub-THz bands [23]. Beneath the patch layer, a prepreg layer Rogers RO4450F ($\varepsilon_r$ = 3.52) is introduced to laminate the multilayer structure. The slot layer is then implemented on a Rogers RO3006 substrate ($\varepsilon_r$ = 6.15), which enables the slot-coupled feeding mechanism between the feeding network and the radiating patches. This multilayer configuration provides efficient EM coupling while maintaining good isolation between different functional layers.

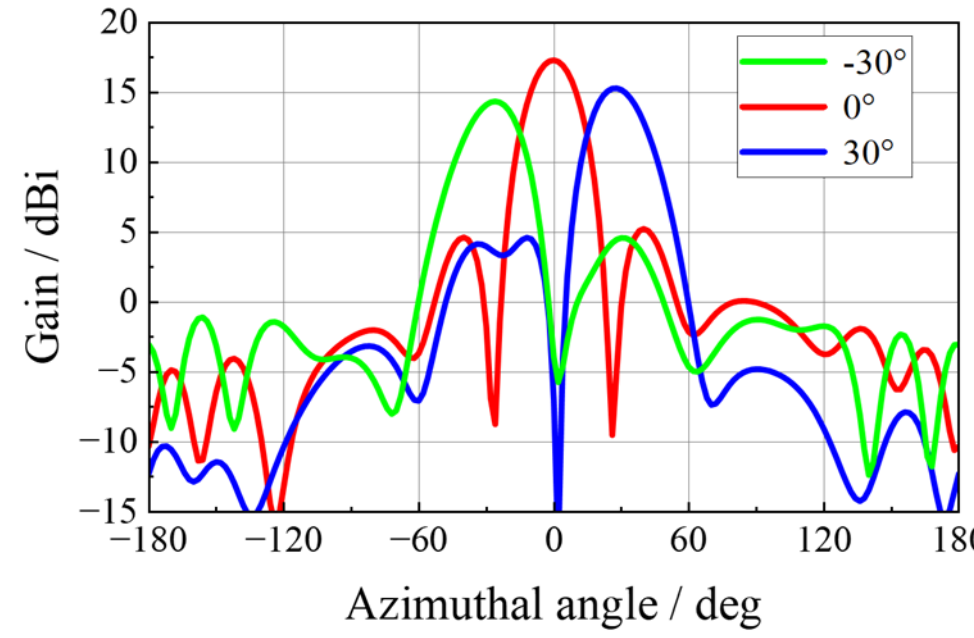

Fig. 7. Simulated radiation patterns of various subarrays for different beamswitching angles.

To integrate the switching devices and the associated biasing network, the feeding network layer is introduced below the slot structure. In this layer, the MASW-011029 switches are mounted as bare dies and electrically interfaced through gold bond wires. The DC biasing lines are routed independently from the RF paths to reduce undesired coupling between the control circuit and the RF signals. Finally, an FR4 support layer and a ground plane are introduced at the bottom of the structure to provide mechanical stability and a common ground reference for the entire RIS. The Rogers RO3006 and FR4 substrates are mechanically assembled using nylon screws, leaving an air gap between them to accommodate the bare dies and bond wires on the feeding network layer. Under this stackup configuration, the complete RIS array layout is constructed as shown in Fig. 8(b). The proposed RIS consists of multiple controllable subarrays arranged across the aperture. Each subarray contains a group of unit elements that share the same switching control, enabling directional beam manipulation with a reduced number of switches. This configuration not only simplifies the wire-bonding process and biasing network routing but also improves the overall scalability of the RIS platform for large-aperture implementations at sub-THz frequencies. More importantly, from a signal-processing and control perspective, the subarray concept can effectively reduce the number of independent RIS control signals. This reduction lowers the control signaling overhead, simplifies the biasing and control interface, and decreases the computational burden associated with RIS state configuration. Therefore, this approach is beneficial for practical large-aperture RIS implementations, where minimizing the control signaling overhead is highly desirable.

Following the multilayer RIS architecture described above, the practical integration of the switching device into the RF feeding network becomes another important design consideration. As discussed in Section II-A, the selected MASW-011029 switch is provided in a bare-die form and requires gold wire bonding to establish both RF and DC connections. At sub-THz frequencies around 100 GHz, the parasitic inductance introduced by the bonding wires can significantly influence the impedance matching and signal routing performance. Therefore, the bonding-wire configuration needs to be carefully designed to maintain stable RF behavior. The bonding-wire implementation used in this work is illustrated in Fig. 9. Two RF switches are mounted on the feeding network layer, and the RF connections between the switch pads and the PCB transmission lines are realized through short gold bond wires. Since each RF pad requires wire bonding to establish the RF path, multiple bond wires are implemented on each pad. This multi-wire bonding configuration effectively reduces the bonding inductance and improves the robustness of the RF interconnection at sub-THz frequencies. In addition to the RF interconnections, separate bond wires are used to route the DC bias signals required for switch control. The RF and DC bonding paths are spatially separated to minimize interference between the control signals and the high-frequency RF paths. Another important consideration is the physical length of the bonding wires. Since the inductance of a bonding wire is approximately proportional to its length, excessive wire length may introduce noticeable impedance mismatch at sub-THz frequencies. Therefore, the switches are positioned close to the RF transmission lines so that the bond wires remain short and nearly vertical. This configuration helps maintain stable RF performance across the operating frequency band while also improving assembly reliability. Overall, the bonding-wire integration plays a critical role in enabling practical PCB-based RIS implementations at sub-THz frequencies. By carefully arranging the RF and DC bond wires and maintaining a compact interconnection structure, the proposed design provides a feasible solution for integrating high-frequency bare-die switches with multilayer PCB RIS architectures.

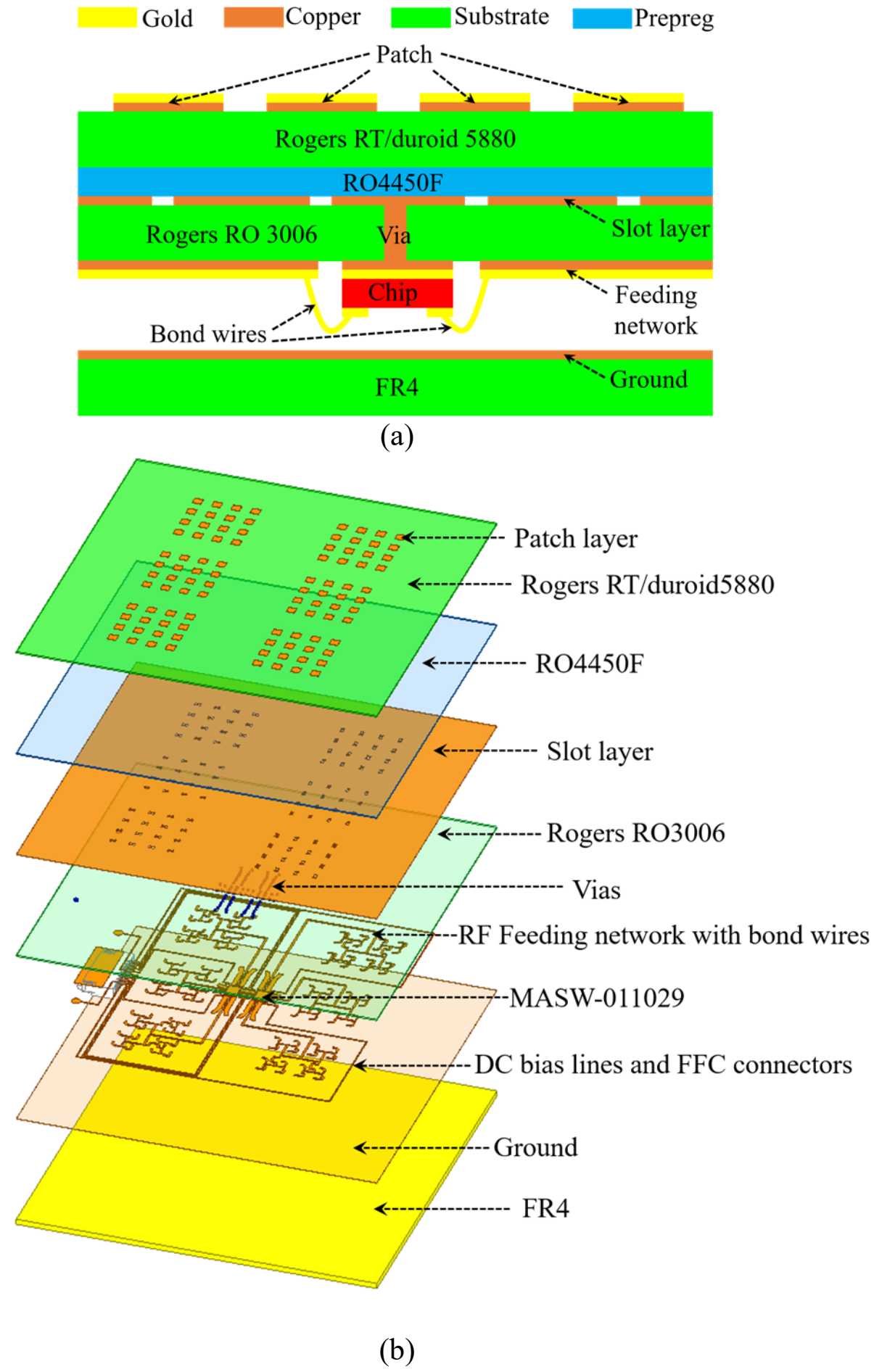

Fig. 8. (a) The RIS stackup. (b) The RIS array configuration.

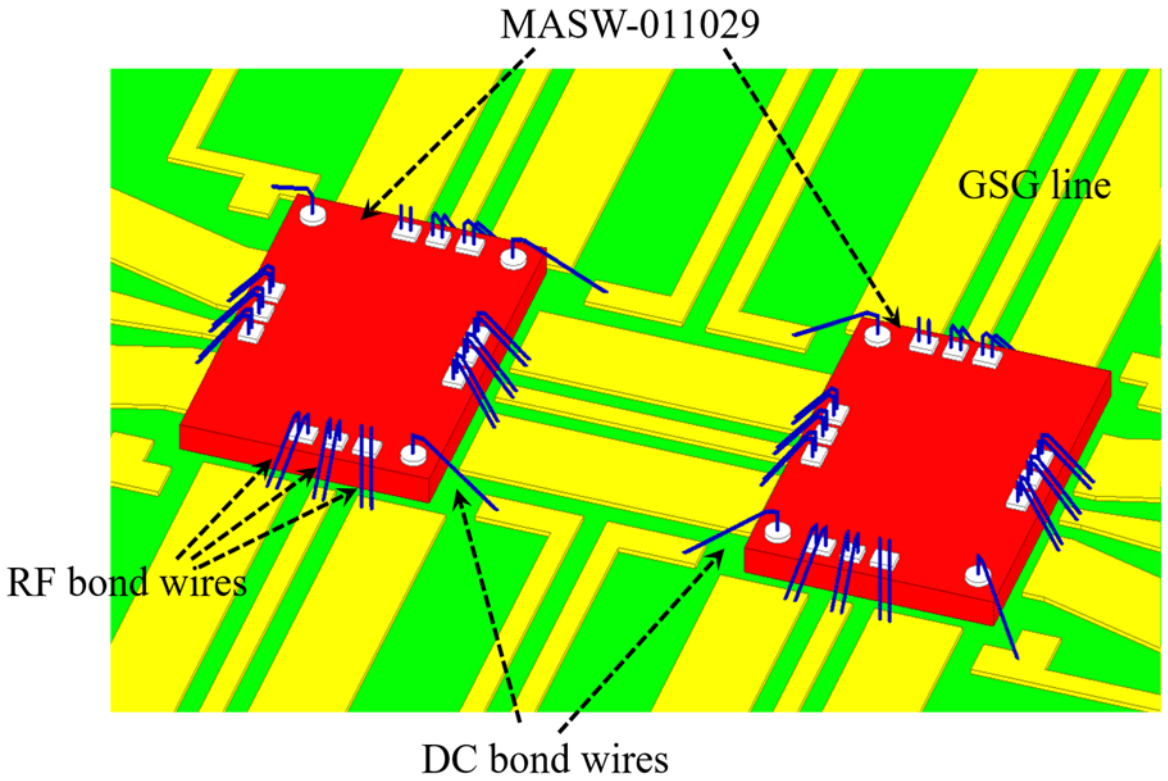

Fig. 9. The bonding wire deign for the RIS.

### *B. Array Simulations and Final Layout*

The geometrical layout and parameter definitions of the proposed RIS are illustrated in Fig. 10, while the corresponding dimensional values are summarized in Table I. As shown in Fig. 10, the figure defines the key structural parameters of the radiating patches, slot coupling structure, feeding network traces, and the bonding interface associated with the switching devices. It is worth noting that the dimensional constraints of the design are closely related to the practical fabrication capability of multilayer PCB. In conventional PCB manufacturing processes, the typical fabrication resolution requires the line width and slot width to be larger than 0.1 mm. In certain special cases, such as single-layer PCB structure, laser fabrication techniques can be employed to achieve significantly higher resolution. For example, gaps as small as 15 μm have been demonstrated using laser-based fabrication, as reported in our previous work [24]. However, such laser fabrication techniques are generally not applicable to multilayer PCB structures due to the alignment requirements between different metallic layers. Since the proposed RIS adopts a multilayer PCB configuration, the entire structure needs to be compatible with standard multilayer PCB fabrication processes. Therefore, the critical geometrical parameters, including the transmission line widths and spacing within the feeding network, are designed around the practical fabrication limit of approximately 0.1 mm. While this constraint ensures reliable fabrication and layer alignment, it also introduces certain performance degradation compared with ideal high-resolution structures. The optimized dimensional values obtained from the design process are listed in Table I, where all parameters are given in millimeters. In addition, the feeding network employed in this design is optimized to realize a beamforming direction of 30°. A similar optimization procedure can be applied to obtain a 0° beamforming configuration. For brevity, the corresponding parameters are not repeated here.

To evaluate the EM performance of the proposed RIS, full-wave simulations are conducted using the configuration illustrated in Fig. 11(a). The simulation setup is consistent with our previous works in [13], [14], where the incident wave arrives at an angle of 30° and the reflected beam is designed toward 0°. In this configuration, a horn antenna illuminates the receiving subarrays of the RIS, and the scattered field distribution is evaluated in the far-field region. Two switching conditions corresponding to the RIS ON and RIS OFF states are considered to examine the reconfigurable behavior of the structure. The simulated 3D radiation patterns presented in Fig. 11 (a) show a clear difference in the scattering distribution of the orthogonally polarized reflected field with respect to the incident wave between the two states, indicating that the integrated switching network effectively modifies the EM response of the RIS aperture. To enable a quantitative comparison, the normalized radiation patterns along the elevation plane are further extracted and plotted in Fig. 11(b). As observed, the RIS ON state produces a pronounced beam toward the desired reflection direction of 0°, whereas the RIS OFF state exhibits a relatively low scattering response. At the designed beam reflection angle of 0°, a gain enhancement of 17.9 dB is achieved compared with the RIS OFF state. These results demonstrate that the proposed RIS architecture can effectively manipulate the reflected wave distribution and realize directional beamforming in the sub-THz band. It is worth noting that the presented simulation results do not include the insertion loss introduced by the RF switches. The impact of the switch loss will be further discussed in the subsequent fabrication and measurement section.

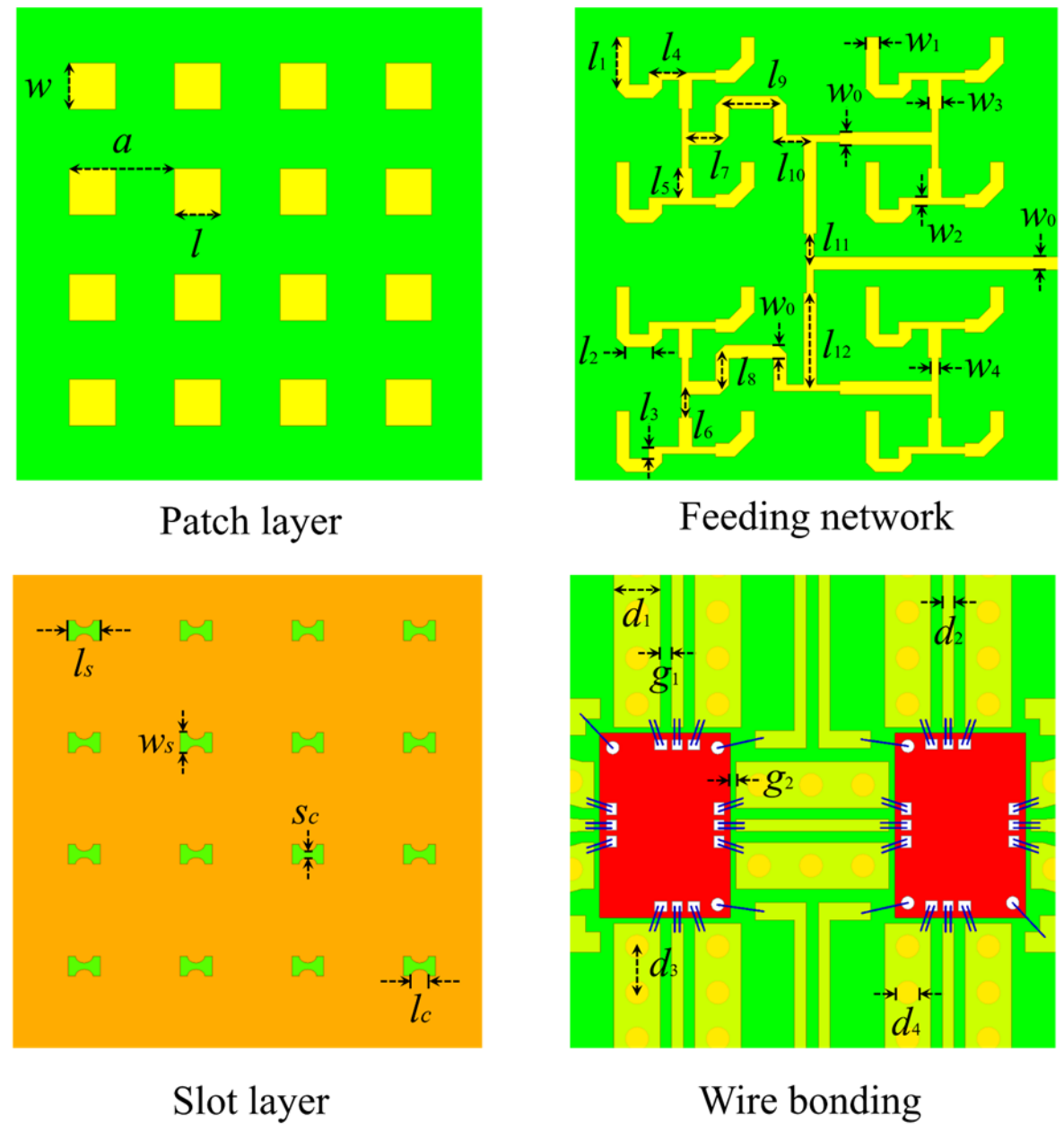

Fig. 10. The RIS dimensions.

TABLE II DIMENSIONS OF THE SUB-THZ RIS DESIGN

| All the variables are in mm | | | | | | | | | |
|---|---|---|---|---|---|---|---|---|---|
| $a$ | $l$ | $w$ | $l_s$ | $w_s$ | $s_c$ | $l_c$ | $l_1$ | $l_2$ | $l_3$ |
| 1.71 | 0.751 | 0.751 | 0.49 | 0.306 | 0.102 | 0.265 | 0.704 | 0.37 | 0.166 |
| $l_4$ | $l_5$ | $l_6$ | $l_7$ | $l_8$ | $l_9$ | $l_{10}$ | $l_{11}$ | $l_{12}$ | $w_0$ |
| 0.495 | 0.397 | 0.408 | 0.51 | 0.497 | 0.797 | 0.498 | 0.408 | 1.302 | 0.18 |
| $w_1$ | $w_2$ | $w_3$ | $w_4$ | $d_1$ | $d_2$ | $d_3$ | $d_4$ | $g_1$ | $g_2$ |
| 0.184 | 0.1 | 0.184 | 0.1 | 0.4 | 0.1 | 0.4 | 0.2 | 0.1 | 0.05 |

### C. Control Circuit

To control the switching states of the proposed RIS, a dedicated biasing circuit is implemented using the MADR-009190 driver [25]. The schematic and layout of the control circuit are shown in Fig. 12(a) and Fig. 12(b), respectively. The driver converts digital logic signals into the required bias voltages and currents to control the MASW-011029 SP3T switch. Each RF path is controlled through the corresponding bias pads (B2, B3, and B4). During operation, the isolation paths require a forward bias current of approximately 10 mA, while the selected RF path is activated using a reverse bias voltage. In the implemented circuit, the supply voltages are configured as $V_{CC}$ = +5 V, $V_{OPT}$ = +5 V, and $V_{EE}$ = −5 V, enabling proper operation of the switching device. Meanwhile, several passive components are incorporated in the control circuit. The bias resistors connected to the switch control pads use a value of 320 Ω, while the coupling capacitors adopt a value of 470 pF. In addition, 0.1 μF decoupling capacitors are placed close to the supply pins to suppress power noise and stabilize the bias voltages. The control board provides multiple channels to independently drive different RIS subarrays and interfaces with external digital controllers for programmable RIS operation.

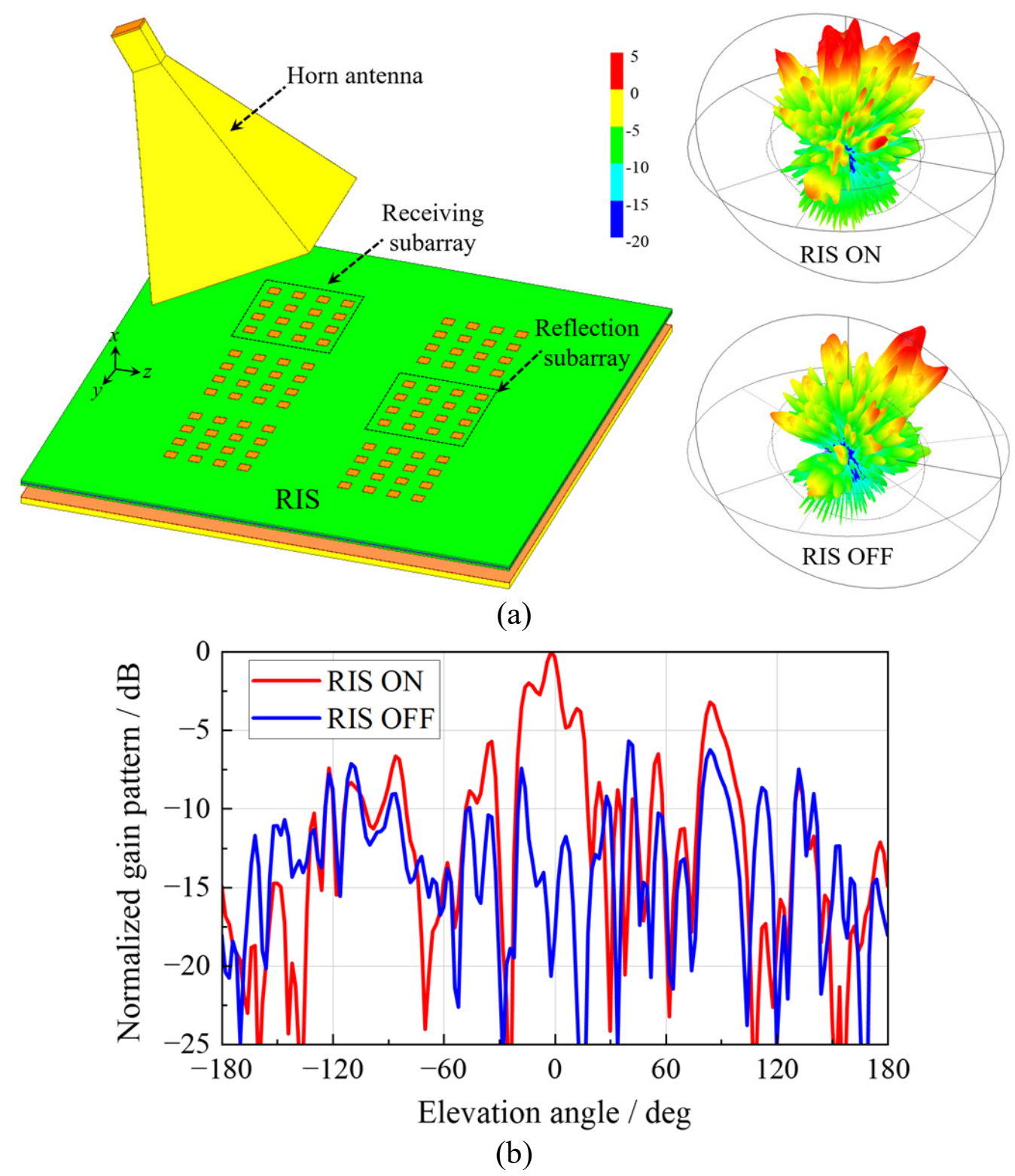

Fig. 11. Full-wave simulation of the RIS. (a) Simulation setup and 3D radiation pattern comparison. (b) 2D normalized radiation pattern comparison between the RIS ON and OFF states.

## IV. FABRICATION AND MEASUREMENT

### A. Fabricated RIS Prototype

The fabricated prototype of the proposed sub-THz RIS is shown in Fig. 13. The RIS structure is implemented using the multilayer PCB stackup described in Section II. As illustrated in Fig. 13(a), the top side of the RIS board contains the radiating patch arrays, while the bottom side integrates the RF feeding network together with the switching components and DC connectors. The top patch layer is fabricated with a gold plating thickness of approximately 0.5 μm. In contrast, the feeding network layer adopts a thicker gold plating of about 2 μm in order to ensure reliable gold wire bonding between the PCB transmission lines and the bare-die RF switches. The integration of the switches is realized through gold wire bonding, as shown in Fig. 13(b). The MASW-011029 switches are mounted as bare dies on the feeding network layer. Since the backside metal of the MASW-011029 die serves as the ground terminal, the switches are attached to the feeding network ground layer using conductive silver epoxy. The feeding network ground layer is further connected to the slot layer through multiple vias, which together form the common DC ground reference of the RIS structure.

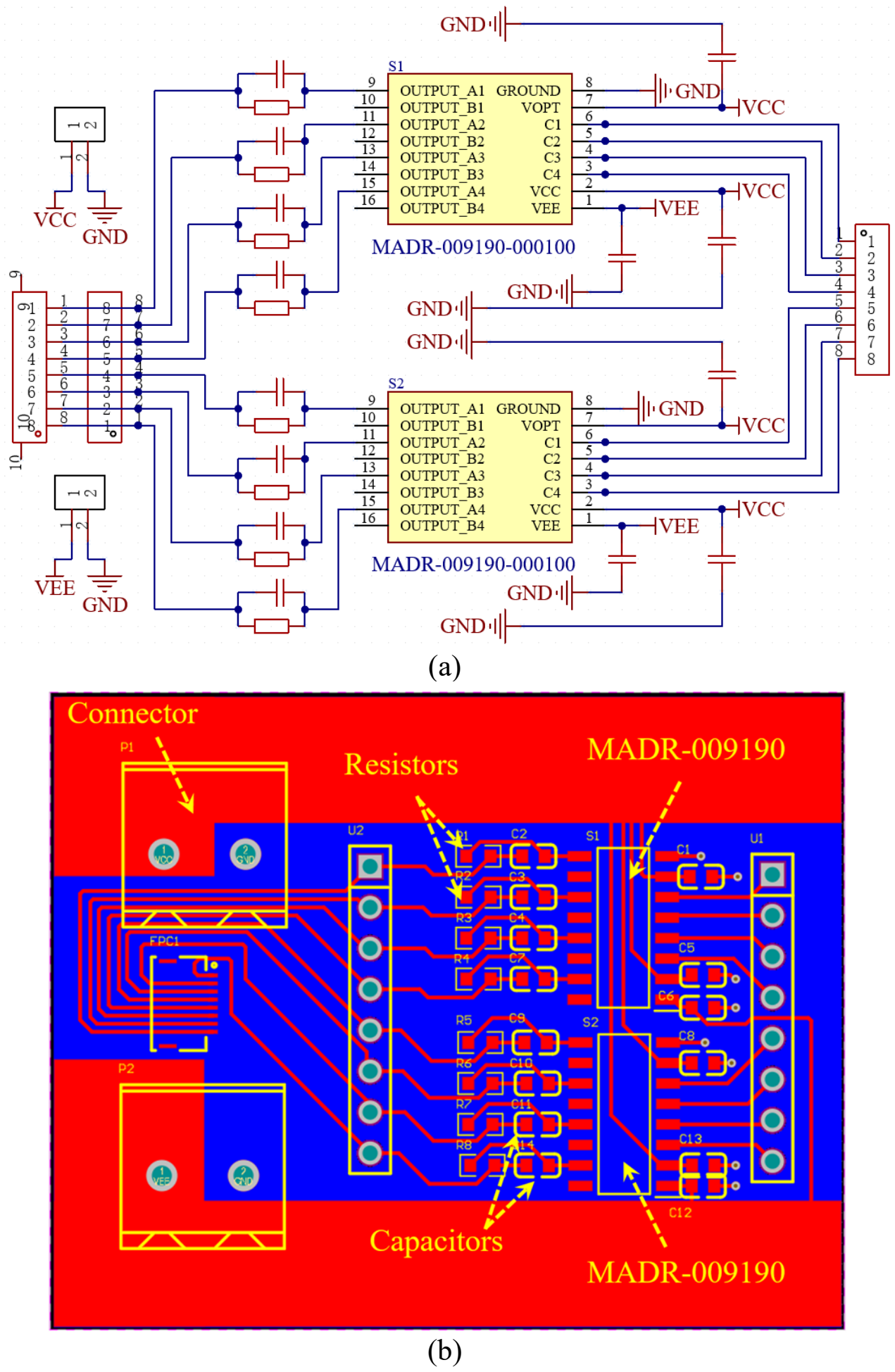

Fig. 12. The designed control circuit for the proposed RIS. (a) Schematic. (b) Layout.

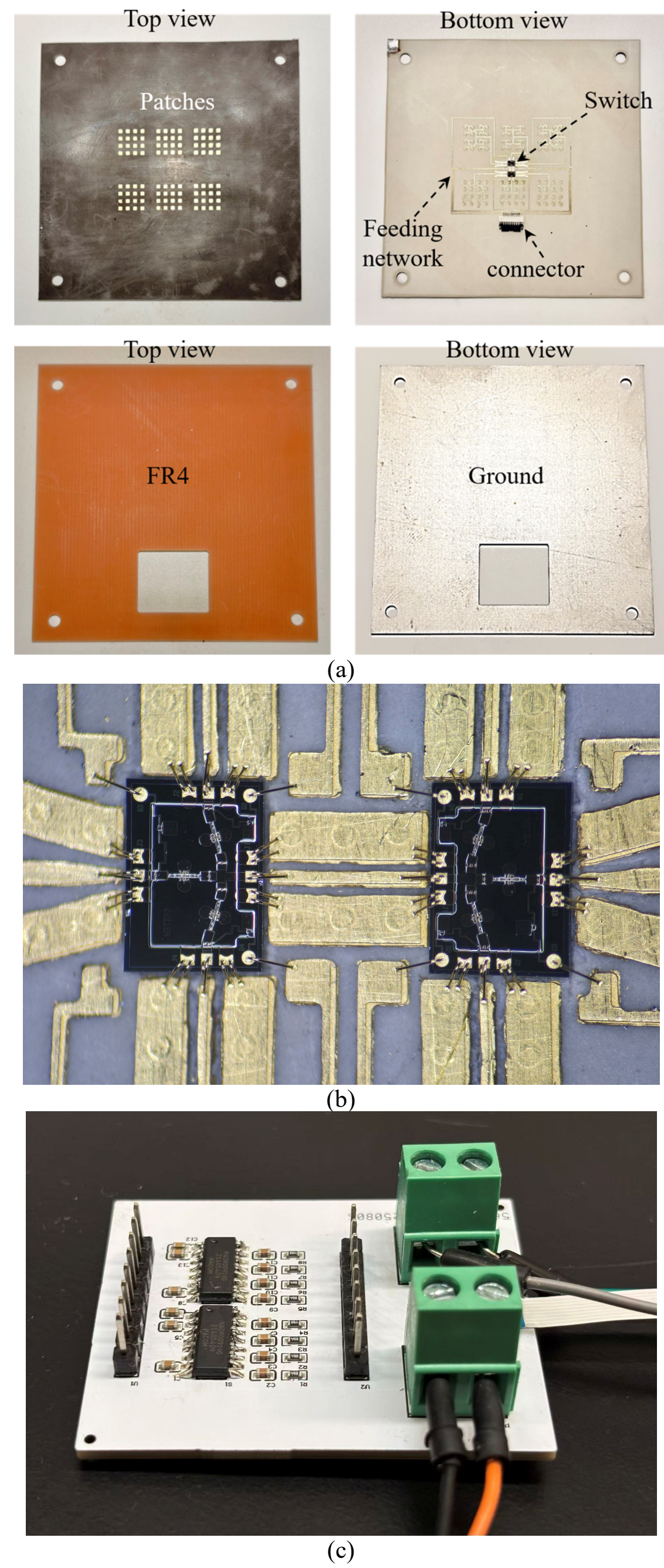

Fig. 13. Fabricated RIS prototypes. (a) RIS and ground boards. (b) bond wires. (c) control circuit.

The RF signal paths between the switch pads and the PCB transmission lines are established using gold bond wires with a diameter of 1 mil. For each RF pad, two bond wires are implemented in parallel in order to reduce the parasitic inductance introduced by the bonding interconnection at sub-THz frequencies. The bonding loop height is approximately 5 mil, while the bonding wire length is around 10 mil, which further helps minimize the inductive parasitics of the RF connection. For the DC bias connections, only a single bond wire is implemented for each DC pad since the DC path does not impose stringent RF parasitic constraints, and a single bonding wire is sufficient for reliable bias delivery. The DC bonding wires exhibit a resistance lower than 0.2 Ω and can support current levels ranging from 0 to around 500 mA, which is sufficient for the bias requirements of the switching devices. The RF and DC bonding paths are spatially separated to reduce coupling between the high-frequency RF signals and the control lines.

To control the switching states of the RIS subarrays, a dedicated control circuit board is fabricated as shown in Fig. 13(c). The control board integrates the MADR-009190 driver together with the required biasing network and connectors. The RF board of the RIS and the DC control circuit are interconnected through a flexible flat cable (FFC), which enables convenient routing of the bias signals while maintaining mechanical flexibility during measurement. In order to accommodate the FFC connector and ensure proper mechanical assembly, a cavity is etched in the ground support board beneath the connector region. This cavity provides sufficient space for the connector and cable routing without disturbing the ground plane structure. In addition, the RIS RF board and the ground board are assembled using nylon screws together with insulating spacers, providing stable mechanical fixation while avoiding undesired electrical interaction with the RF structure.

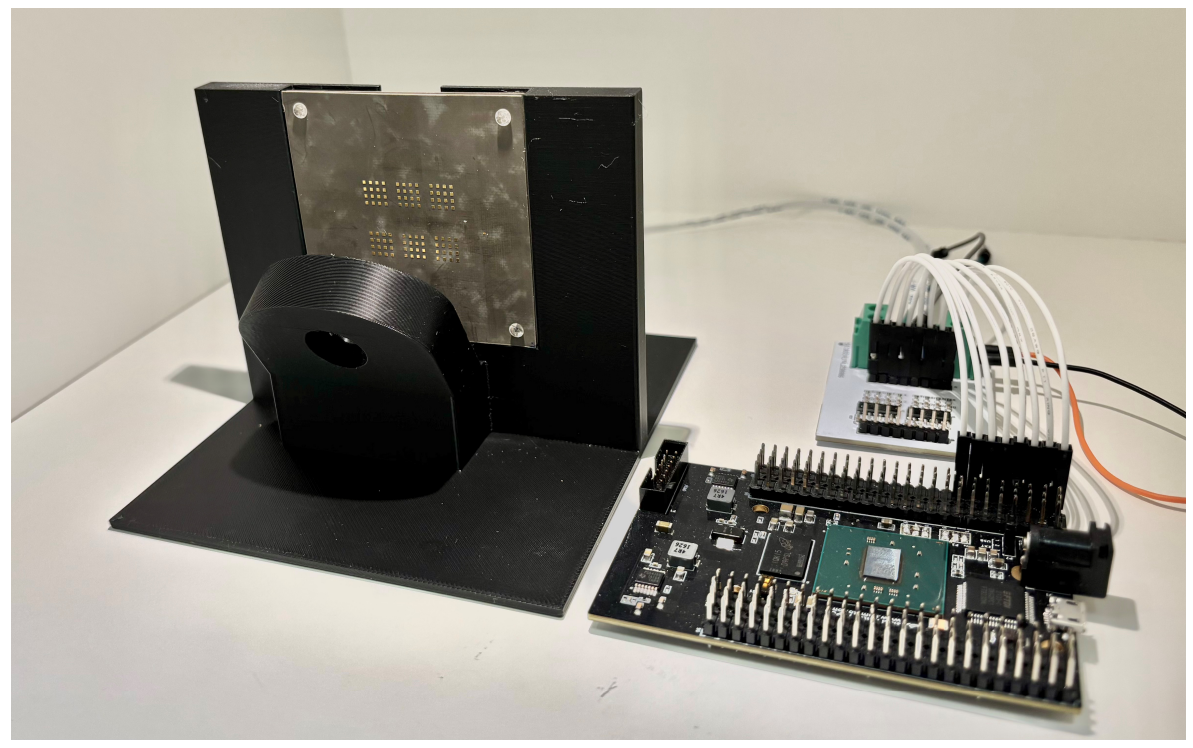

Fig. 14.The entire RIS prototype.

The complete RIS prototype assembled for measurement is shown in Fig. 14. The prototype consists of the RIS RF board, the ground support board, the control circuit board, and the digital control platform. The RIS RF board is vertically mounted on the measurement fixture using a customized mechanical support structure fabricated using a Bambu Lab H2D 3D printer with PLA filaments. This 3D-printed support structure provides stable mechanical fixation and ensures accurate alignment between the incident horn antenna and the RIS aperture during measurement. The switching states of the RIS are controlled through the dedicated biasing circuit described in Section Ⅲ-C. A Skoll Kintex-7 FPGA module is employed to generate the digital control signals required by the MADR-009190 driver, which in turn provides the corresponding bias voltages for the MASW-011029 switches. The FPGA module is connected to a computer through a high-speed communication interface, enabling programmable control of the RIS states in real time. This configuration allows flexible switching control with high switching speed and also provides the capability to implement time-modulated RIS operation when required.

## *B. RIS Measurement*

Before measuring the RIS performance, the RF characteristics of the switch need to be characterized first. The measured transmission and isolation performance of the MASW-011029 switch are presented in Fig. 15. As observed in Fig. 15, the switch exhibits relatively low insertion loss within the lower portion of the operating band. At 100 GHz, the measured transmission parameter is approximately −3.4 dB. However, when the frequency increases toward the upper end of the band, the transmission loss gradually increases and reaches approximately −8.1 dB at 110 GHz. This behavior is mainly attributed to the increased conductor loss and parasitic effects associated with the switch and bonding interconnections at sub-THz frequencies. As a result, the RIS performance above

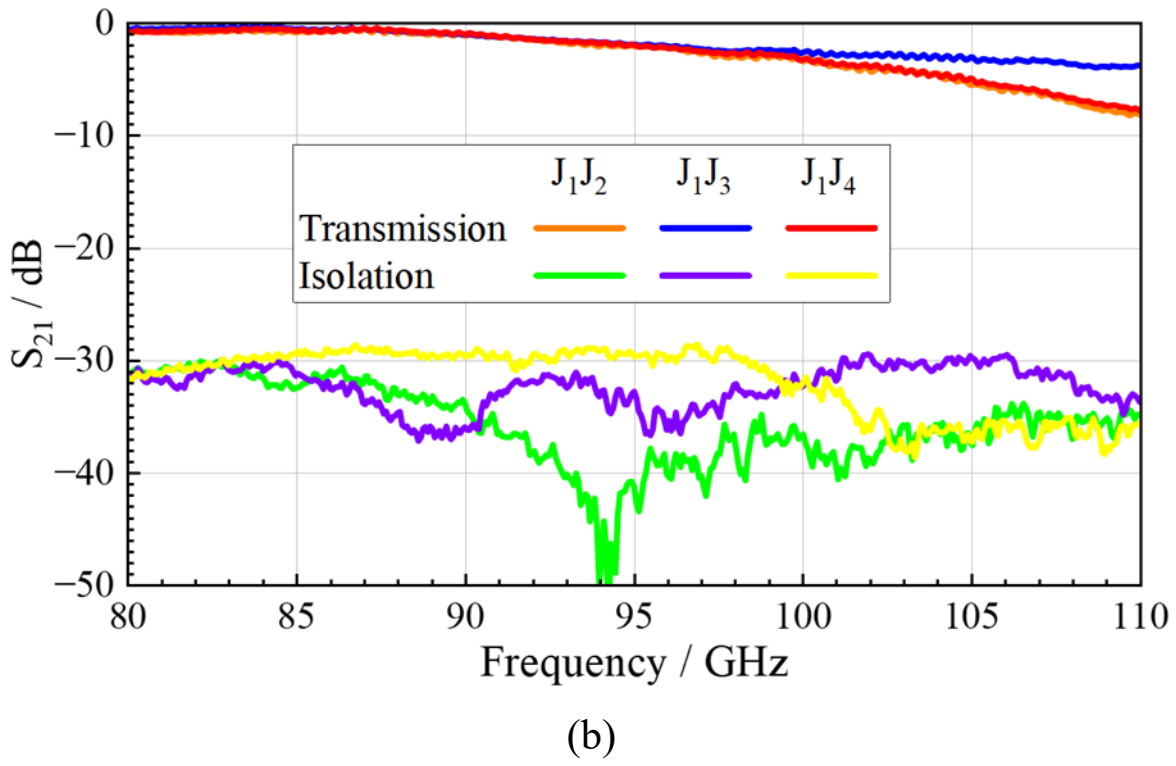

(b)

Fig. 15. Measured transmission and isolation performance of the MASW-011029 switch.

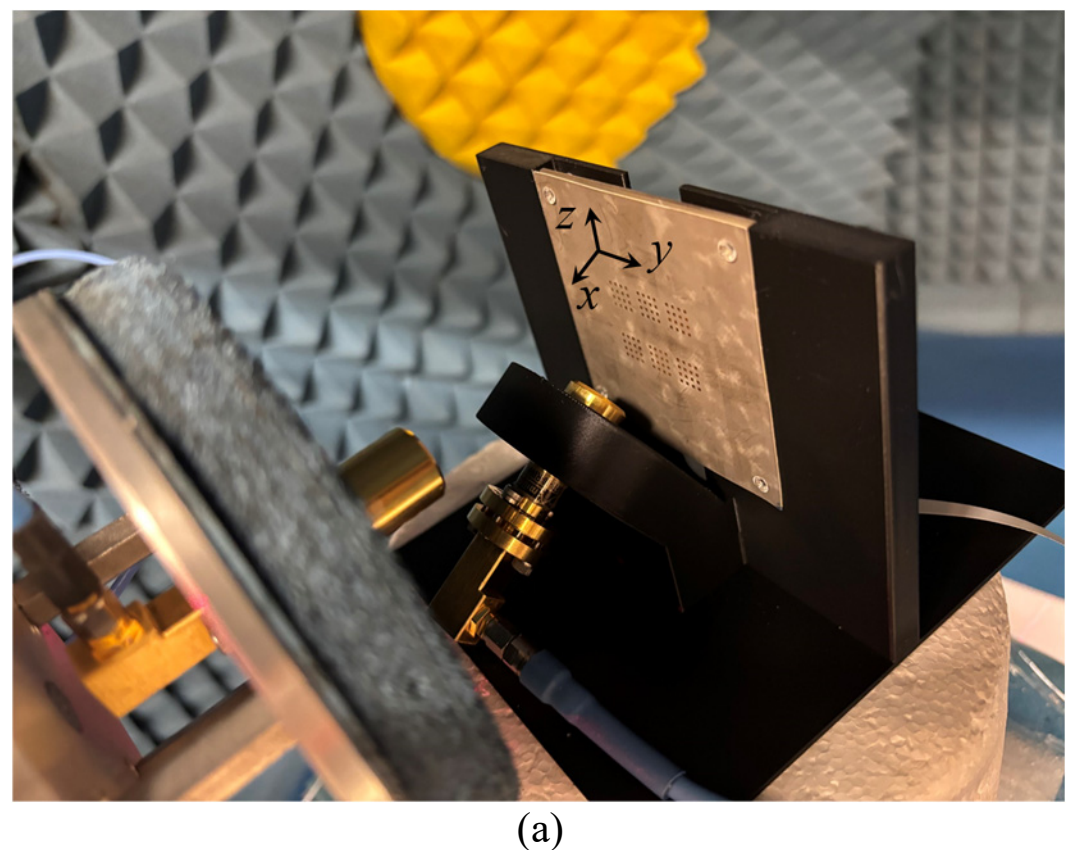

(a)

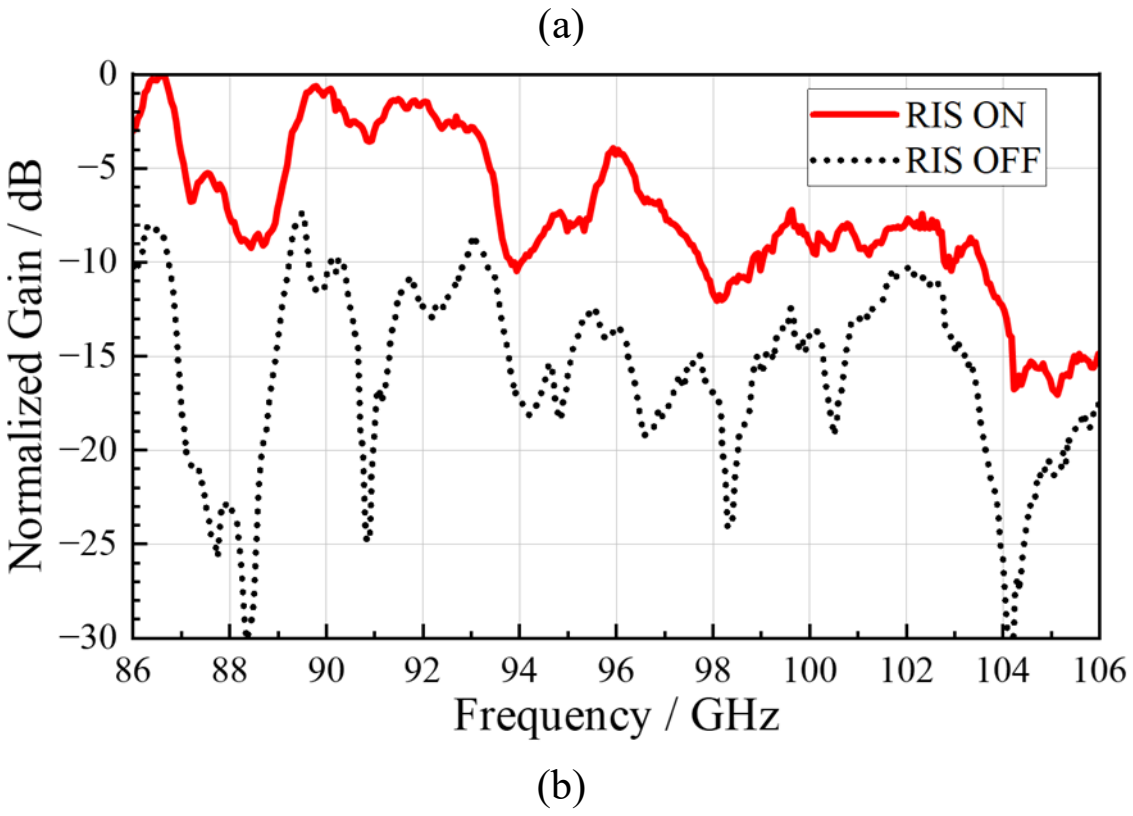

(b)

Fig. 16. (a) Measurement setup for scenario 1 with an incidence angle of $\theta$ = 120°; $\varphi$ = 0° and a reflection angle of $\theta$ = 90°; $\varphi$ = 0°. (b) Measured results.

100 GHz tends to suffer from higher switch loss. Meanwhile, the switch demonstrates excellent isolation performance across the entire measured band. The isolation between different RF paths remains better than −26 dB. Such high isolation effectively suppresses undesired signal leakage between different RF paths and ensures that the selected switching state dominates the EM response of the RIS.

Following the switch characterization, the gain enhancement of the fabricated RIS prototype is experimentally evaluated. The measurement setups for two representative scenarios are presented in Fig. 16 and Fig. 17, respectively. In both cases, a standard-gain horn antenna SFH-10-R0000 is used to illuminate the RIS aperture, while another horn antenna is positioned at the designed reflection direction to capture the scattered field. The distance between the receiving horn antenna and the RIS is selected with 60 mm to satisfy the far-field condition of each subarray for the operating frequency band. Fig. 16(a) illustrates the measurement configuration for Scenario 1, where the incident wave arrives at an angle of θ = 120°, φ = 0°, and the reflected beam is measured at θ = 90°, φ = 0°. The corresponding measured results are plotted in Fig. 16(b). The RIS ON state produces a strong reflection toward the desired direction, while the RIS OFF state exhibits a significantly weaker scattered field. As observed from the measured results, the proposed RIS achieves a noticeable gain enhancement across the entire operating band. Specifically, within the frequency range of 86–100 GHz, the measured gain enhancement remains close to 10 dB, indicating effective beam redirection by the RIS. When the frequency increases to the range of 100–106 GHz, the gain enhancement gradually decreases to approximately 5 dB, mainly due to the increased RF loss introduced by the switching devices at higher frequencies. Overall, the measured gain enhancement reaches an average value of 8.4 dB across the frequency range from 86 GHz to 106 GHz, demonstrating the broadband beamforming capability of the proposed RIS.

To further validate the beam manipulation performance under a different reflection direction, an additional measurement scenario is conducted, as shown in Fig. 17(a). In this configuration, the incident angle remains θ = 120°, φ = 0°, while the reflected beam is measured at θ = 90°, φ = 30°. The corresponding measurement results are presented in Fig. 17(b). Similar to scenario 1, a clear difference between the RIS ON and RIS OFF states can be observed. The RIS ON state produces a strong scattered signal toward the desired reflection direction, confirming the reconfigurable beam steering capability of the RIS. Across the frequency band of 86–100 GHz, the gain enhancement again remains close to 10 dB, while within 100–106 GHz the enhancement is approximately 5 dB. The average gain enhancement obtained in scenario 2 is approximately 9.2 dB over the entire 86–106 GHz frequency band, which is consistent with the results obtained in Scenario 1.

The analysis of the measured results is shown below. The simulation results reported in Fig. 10 do not include the RF losses introduced by the switches and the bonding interconnections. To better understand the discrepancy between simulation and measurement, a post-simulation analysis is performed by incorporating the measured switch loss together with the parasitic effects of the bonding wires. At 100 GHz, the combined loss contributed by the switch and the wire-bonding interconnection is estimated to be approximately 5.9 dB for each RF path. Considering that the RIS architecture involves two such RF paths along the signal routing, the total additional loss introduced by these components becomes approximately 11.8 dB. When this loss is subtracted from the ideal simulated gain enhancement of 17.9 dB, the expected gain enhancement reduces to approximately 6 dB, which is in good agreement with the measured gain enhancement of about 5 dB at frequencies around and above 100 GHz. In addition to the switch insertion loss, several practical factors contribute to the observed performance degradation, including the parasitic inductance of the gold bond wires and the increased conductor loss caused by PCB surface roughness at sub-THz frequencies. Fabrication tolerances of the multilayer PCB process together with minor measurement misalignments may further introduce deviations between the simulated and measured results. The power consumption of the RIS prototype is also evaluated during the measurement. In the implemented design, the measured DC supply is 5 V with a current of approximately 33 mA, corresponding to a total power consumption of about 0.165 W. This relatively low power consumption demonstrates the advantage of the subarray-based architecture, which significantly reduces the number of required switching devices compared with unit-level control schemes.

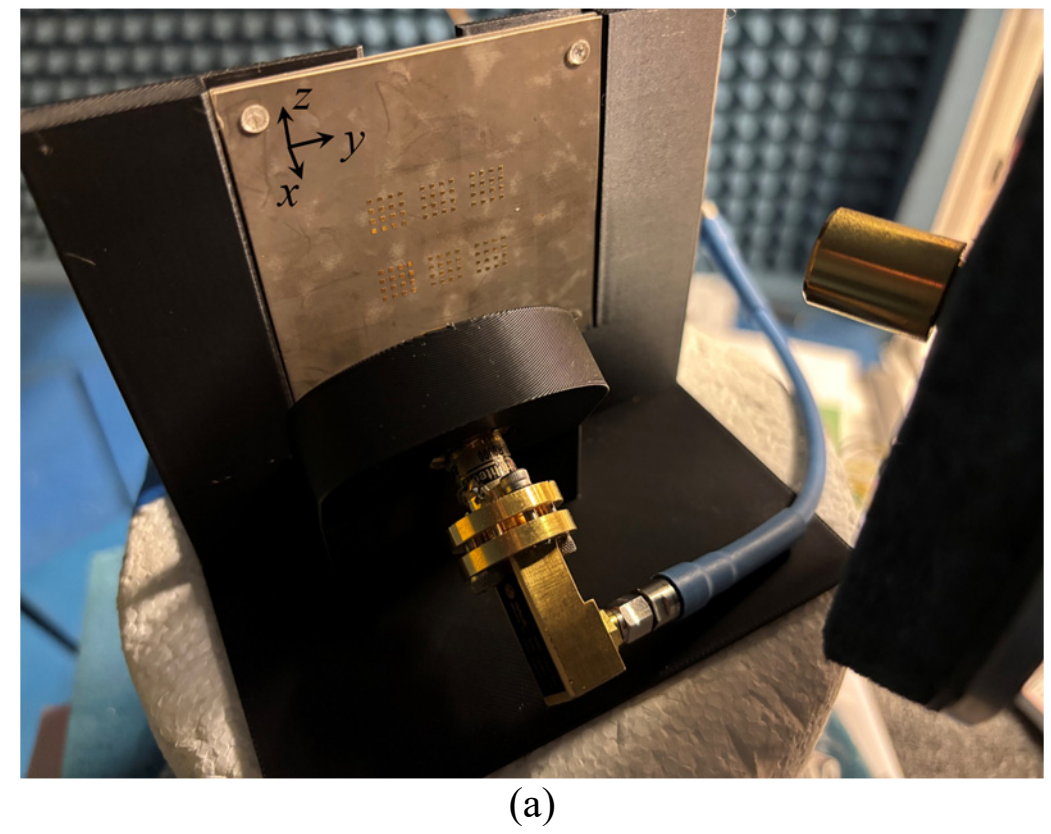

(a)

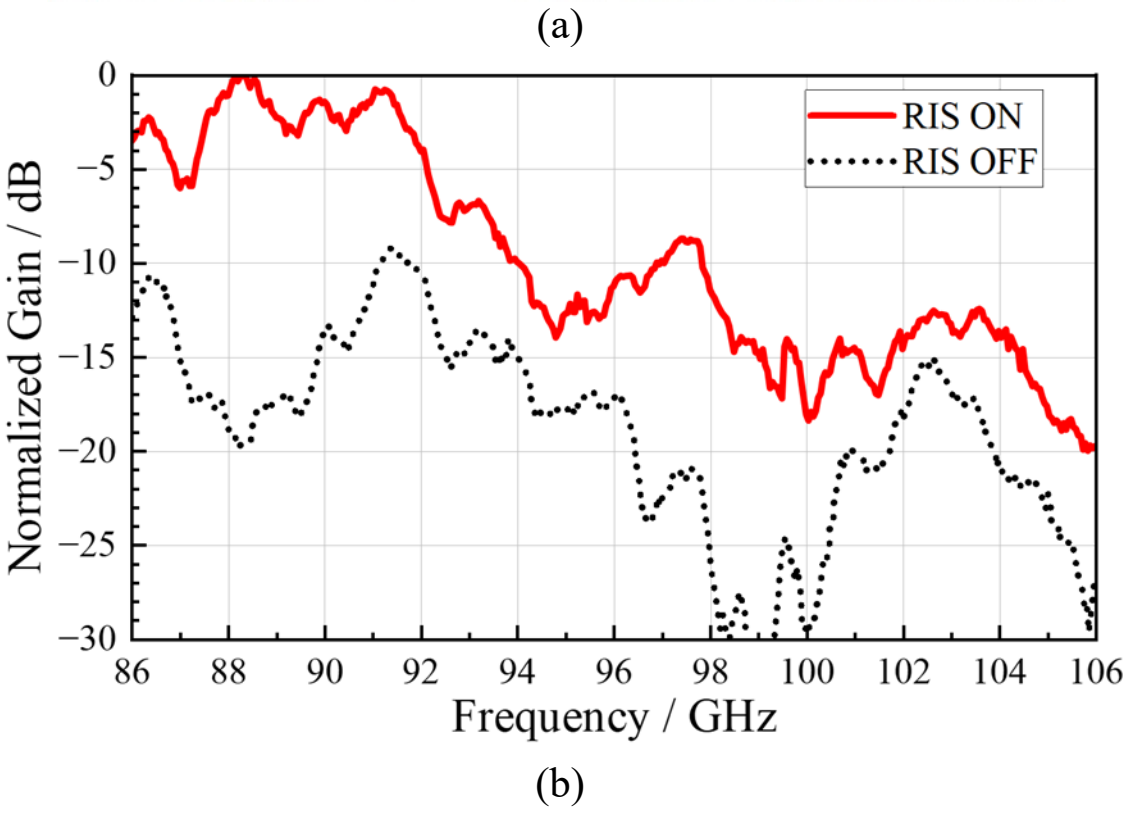

(b)

Fig. 17. (a) Measurement setup for scenario 2 with an incidence angle of $\theta$ = 120°; $\varphi$ = 0° and a reflection angle of $\theta$ = 90°; $\varphi$ = 30°. (b) Measured results.

Despite these practical loss mechanisms, the experimental results validate that the proposed RIS can effectively manipulate the reflected EM waves and achieve consistent gain enhancement across a wide frequency range around 100 GHz. The measured performance verifies the feasibility of implementing wideband RIS using multilayer PCB technology combined with high-speed AlGaAs switches in the sub-THz band.

## V. Conclusion

This work presents a wideband multilayer PCB-based RIS operating in the sub-THz band. The proposed RIS integrates slot-coupled patch elements with AlGaAs SP3T switches and adopts a subarray-based architecture to reduce hardware complexity while enabling fast beam reconfiguration. A 12 × 8 prototype is fabricated and experimentally characterized. The measured results show an average gain enhancement of about 8.4 dB across 86–106 GHz, validating the feasibility of implementing wideband RIS platforms around 100 GHz using PCB technology. Future work will focus on reducing RF loss associated with sub-THz switching devices and interconnections, extending the RIS to larger arrays with finer control resolution, and exploring dynamic beam control schemes such as time-modulated RIS operation.